\documentclass[creativecommons]{eptcs}
\providecommand{\event}{GandALF 2021}
\usepackage{underscore}           
\usepackage[T1]{fontenc}
\usepackage[utf8]{inputenc}
\usepackage{times}
\usepackage{algorithm}
\usepackage{algorithmic}
\usepackage{amsfonts}
\usepackage{amsmath}
\usepackage{amssymb}
\usepackage{babel}
\usepackage{graphicx}
\usepackage{hyperref}
\usepackage{nameref}
\usepackage{subcaption}
\usepackage{subfiles}
\usepackage{syntax}
\usepackage{tikz}
\usepackage{tipa}
\usepackage{todonotes}
\usepackage{wrapfig}
\usepackage{xcolor}
\usepackage{color, colortbl}
\definecolor{Gray}{gray}{0.9}

\def\titlerunning{Decentralized LTL Enforcement}
\def\authorrunning{F. Gallay \& Y. Falcone}

\newtheorem{definition}{Definition}
\newtheorem{example}{Example}
\newtheorem{property}{Property}
\newtheorem{lemma}{Lemma}

\newtheorem{corollary}{Corollary}

\definecolor{dark-gray}{gray}{0.5}
\newcommand{\secref}[1]{Sec.~\ref{#1}}
\newcommand{\algoref}[1]{Algorithm~\ref{#1}}
\newcommand{\lemmaref}[1]{Lemma~\ref{#1}}
\newcommand{\LTL}{$\mathit{LTL}$}
\newcommand{\LTLm}{\mathit{LTL}}
\newcommand{\LTLp}{\LTLm_\mathrm{p}}
\newcommand{\LTLf}{\LTLm_\mathrm{f}}
\newcommand{\Next}{\textsc{\textbf{X}}}
\newcommand{\Globally}{\textsc{\textbf{G}}}
\newcommand{\Eventually}{\textsc{\textbf{F}}}
\newcommand{\Until}{\textsc{\textbf{U}}}
\newcommand{\Release}{\textsc{\textbf{R}}}

\newcommand{\apFormula}{\texttt{apFormula}}
\newcommand{\apr}{\texttt{apr}}
\newcommand{\rwT}{\texttt{rwT}}

\newcommand{\DNF}{\texttt{DNF}}

\newcommand{\encode}{\texttt{encode}}
\newcommand{\tcl}{\texttt{tcl}}
\newcommand{\TCL}{\texttt{TCL}}
\newcommand{\rwi}{\texttt{rw$_i$}}
\newcommand{\distance}{\texttt{distance}}

\newcommand{\updateTCL}{\texttt{updateTCL}}
\newcommand{\reduction}{\texttt{reduce}}

\newcommand{\gOR}{\ |\ } 
\newcommand{\AP}[0]{\mathit{AP}}

\title{Decentralized LTL Enforcement}
\author{Florian Gallay $\quad$ Yli\`es Falcone
\institute{Univ. Grenoble Alpes, CNRS, Inria, Grenoble INP, Laboratoire d'Informatique de Grenoble, 38000 Grenoble, France}
\email{florian.gallay1@etu.univ-grenoble-alpes.fr, ylies.falcone@univ-grenoble-alpes.fr}
}
\begin{document}
\maketitle
\vspace{-0.5em}
\begin{abstract}
We consider the runtime enforcement of Linear-time Temporal Logic formulas on decentralized systems with no central observation point nor authority.
A so-called enforcer is attached to each system component and observes its local trace.
Should the global trace violate the specification, the enforcers coordinate to correct their local traces.
We formalize the decentralized runtime enforcement problem and define the expected properties of enforcers, namely soundness, transparency and optimality.
We present two enforcement algorithms.
In the first one, the enforcers explore all possible local modifications to find the best global correction.
Although this guarantees an optimal correction, it forces the system to synchronize and is more costly, computation and communication wise.
In the second one, each enforcer makes a local correction before communicating.
The reduced cost of this version comes at the price of the optimality of the enforcer corrections.
\end{abstract}
\section{Introduction}
%
%
Runtime verification~\cite{series/lncs/10457,FalconeKRT21} is the collection of theories, techniques, and tools dedicated to the verification of system executions against a formal specification.
\emph{Runtime enforcement}~(cf.~\cite{FalconeMRS18,FalconeP19}) extends runtime verification and consists in using runtime \emph{enforcers} to \emph{ensure} the absence of violation to the specification. 
The specification is formalized for instance as a Linear-time Temporal Logic (LTL) formula~\cite{Pnueli;1977}.
Usually, the system is seen as a black box; only its execution is observable (not its implementation).
The execution is abstracted as a sequence of events where each event contains the set of relevant atomic propositions that hold on the system state.
In (centralized) enforcement, the sequence of events, called trace, is fed to one (central) enforcer which transforms it and outputs a sequence that does not violate the property.
Usually, enforcers must be \emph{sound}, \emph{transparent} and \emph{optimal}, that is, their output trace should not violate the property, they should only alter the execution if needed (i.e. to prevent property violations), and the alteration should be minimal, respectively.
\vspace{-1em}
\paragraph{Motivation and challenges.}
We consider decentralized systems, that is systems with no central obser\-vation point nor authority but which are instead composed of several components, each producing a local trace.
Decentralized systems abound (e.g., multithreaded processors, drone swarms, decentralized finance), some of which can have safety critical properties to be ensured.
It is desirable to define enforcement techniques for decentralized systems so as to ensure their desired properties.
In the decentra\-li\-zed setting, enforcers should coordinate and modify their local traces in such a way that the global trace respects the enforced property.
\vspace{-1em}
\paragraph{Approach overview.}
We address the problem of enforcing LTL formulas on decentralized systems.
We start by defining the runtime enforcement problem in the decentralized setting.
Then, we define our enforcement algorithms, which intuitively proceed as follows.
Upon each new event $\sigma$ emitted by the system, using LTL expansion laws~\cite{Pnueli;1977}, we transform the formula to be enforced at the current timestamp into what we refer to as a \emph{temporal disjunctive normal form}, where each disjunct is composed of a present and future obligation formula separated.
The enforcers can then evaluate the present obligations and alter their local observation if needed, i.e. when outputting $\sigma$ would violate it.
We note that our enforcers only evaluate the present obligations in the disjuncts.
After their evaluations with the current event, a subset of the disjuncts will have their present obligations different from $\bot$.
Only the future obligations of these disjuncts are kept for the next timestamp.
This strategy spares the rewriting of the future obligations of the discarded disjuncts.
%
When $\sigma$ needs to be corrected, the enforcers keep track of possible corrections of $\sigma$ and of the associated formulas.
As the system is decentralized, each enforcer can only observe some atomic propositions of the system.
Therefore, the enforcers update the associated formulas with their local observations.
Then, they send it to another enforcer that will, in turn, do the same until the present obligations are entirely evaluated.
In our first algorithm, the formula is evaluated with every possible event over the set of atomic propositions of the system.
This allows the enforcers to find the best possible correction if needed.
Through communication, they will naturally build the entire set of possible events (the update can be seen as the exploration of a tree whose leaves represent every possible assignment of the atomic propositions).
After each local update, each enforcer garbage collects the events that cannot be extended to a viable correction of $\sigma$.
Then, once the formula has been entirely evaluated, the output event is so that soundness, transparency, and optimality are preserved.
Our second algorithm proceeds similarly: the only difference is that, instead of exploring the entire set of possible events, each enforcer makes a local decision before sending the function to another enforcer.
This decision consists in choosing one event from the domain (w.r.t. soundness and transparency as well) and only sending this one in order to prevent the exponential growth of the domain.
In both cases, the formula to be enforced in the next timestamp is built after choosing the output event.
Since enforcers make (optimal) local decisions, optimality of the global event is not guaranteed by the second algorithm using the future obligations associated with the emitted event.
\vspace{-1em}
\paragraph{Related work.}
This paper is at the intersection of two topics, namely \emph{decentralized monitoring} and \emph{runtime enforcement}.
In decentralized monitoring (cf.~\cite{FrancalanzaPS18} for an overview), a lot of work has been done on the verification of decentralized systems for several specifications languages such as LTL and finite-state automata.
These research efforts differ mainly in the assumptions they make on the underlying system.
Similarly to this paper, some existing approaches to decentralized monitoring (e.g., \cite{journals/fmsd/0002F16,Colombo2016}) are based on formula rewriting~\cite{Rosu2005}; the specification is usually represented as an LTL formula or in an extension of LTL like MTL~\cite{ThatiR05} and then rewritten and simplified until a verdict can be emitted.
Other approaches focus on monitoring distributed systems and tackle the problem of global predicate detection~\cite{NatarajanCMG17,Vinit2007} or of fault tolerance for monitors~\cite{FaultTolerant2016}, that is, reaching consensus with monitors that are subject to faults.
The aforementioned work performs decentralized monitoring on \emph{centralized} specifications.
In~\cite{El-HokayemF17,El-HokayemF20}, the focus is on monitoring decentralized specifications that is, multiple  interde\-pendent specifications that apply to separate parts of the system.

The above approaches are dedicated to verification in that they focus on determining a verdict but do not consider at all 1) what should be done when the property is violated and 2) what can be done to prevent violations.
Runtime enforcement (cf.~\cite{FalconeMRS18}) approaches try to prevent violations during the execution of the system.
This topic has also seen a lot of research efforts on modeling and synthesizing enforcement monitors from several specifications formalisms for discrete-time~\cite{fmsd/FalconeMFR11,FalconeFM12,DolzhenkoLS15} and timed properties~\cite{fmsd/PinisettyFJMRN14,FalconeJMP16,FalconeP19} and even stochastic systems~\cite{KonighoferRPTB21}.
To the best of our knowledge, the only enforcement approaches for decentralized systems are~\cite{HalleKBEF18,HuDYSD20}, respectively tailored to artifact documents and robotic swarms.
However, in this paper, we formally define the decentralized runtime enforcement problem in a generic manner and provide two generic algorithms.
Finally, we note that the setting of our approach also differs from the one in runtime enforcement techniques with (uncontrollable) actions/events~\cite{BasinJKZ13,RenardFRPJM15,KhouryH15,RenardRF20} where system action/events are blocked/buffered in that our monitors instead directly modify the truth value of some atomic propositions of interest.
\paragraph{Outline.}
\secref{prelimnotions} defines preliminary notions.
\secref{sec:re} introduces the \emph{decentralized runtime enforce\-ment} problem.
\secref{normalizing}, defines the transformation of the formula allowing the separation between the \emph{present} and \emph{future}.
\secref{encoding} defines the data structure used to encode the enforcer state.
In \secref{evaluation}, we define how the enforcers evaluate the formula.
\secref{algoI} and~\ref{algoII} present and compare the algorithms based on global and local exploration, respectively.
\secref{sec:conclusion} concludes and outlines some research avenues.

The extended version~\cite{journals/corr/abs-2107-06084} of this paper contains details about some formula transformations, a complete example of an execution of both algorithms, and proposition proofs.
%

%
\section{Preliminary Notions}
\label{prelimnotions}
%
%
This section introduces some preliminary notions and states the assumptions of our approach.
\paragraph{Decentralized systems.}
A decentralized system consists of $n$ components $C_1, \ldots, C_{n}$.
On each compo\-nent $C_i$, $i \in [1 \ldots n]$, we assume a local set of atomic propositions of interest $\AP_i$.
We also assume that $\{\mathit{AP}_1, \ldots, \mathit{AP}_n \}$ forms a partition of $\mathit{AP}$.
\paragraph{Events and traces.}
An \emph{event} is a set of atomic propositions describing the system state.
For an event $\sigma \in 2^{AP}$, when $p \in \sigma$, it means that atomic proposition $p$ holds on the system.
We denote the set of all events $2^{AP}$ as $\Sigma$ and we call this set the \emph{alphabet}.
Similarly, $\Sigma_i = 2^{AP_i}$ denotes the set of local events to component $C_i$.
Note that $\Sigma \neq \bigcup_{i \in [1,n]} \Sigma_i$.

At runtime, each component $C_i$ emits a \emph{local trace} $u_i$ of events from its local set of atomic propositions $\mathit{AP}_i$.
At any timestamp $t$, the local trace is of the form $u_i(1) \cdot u_i(2) \dotsb u_i(t)$ with $\forall t' < t, u_i(t') \in \Sigma_i$ and where $u_i(j)$ represents the $j$-th local event of $C_i$.
The \emph{global trace} represents the sequence of events emitted by the system as a whole.
At any timestamp $t$, the global trace is of the form: $u = u(1) \cdot u(2) \dotsb u(t)$ with $\forall t' < t, u(t') \in \Sigma$ and where $u(j)$ represents the $j$-th event emitted by the system.
It is possible to build the global trace from the local traces: $u = u_1(1) \cup \dotsb \cup u_n(1) \cdot u_1(2) \cup \dotsb \cup u_n(2) \dotsb u_1(t) \cup \dotsb \cup u_n(t)$ as well as the local traces from the global trace: $u_i = u(1) \cap AP_i \dotsb u(t) \cap AP_i$.
The set of all finite traces over an alphabet $\Sigma$ is denoted as $\Sigma^*$ whereas the set of all infinite traces over $\Sigma$ is denoted as $\Sigma^\omega$. 
The suffix of a (finite or infinite) trace starting at time t is $w^t = w(t) \cdot w(t+1) \dotsb$.
The set of all traces is denoted as $\Sigma^\infty = \Sigma^* \cup \Sigma^\omega$.

To measure the differences between two events, we use a distance function that returns the number of atomic propositions with different value between them.
\begin{definition}[Distance between events]
	Let $\sigma, \sigma'' \in \Sigma$.
	Function \distance $:  \Sigma \times \Sigma \times 2^{\AP} \rightarrow \mathbb{N}$ is defined as follows:
	$ \distance(\sigma, \sigma'', AP) = \#\{AP \cap  (\sigma \cap \overline{\sigma''} \cup \overline{\sigma} \cap \sigma'')\} $, where the complementary events are taken w.r.t. $\AP$.
\end{definition}
Note that we cannot directly use the Hamming Distance because the events are not represented as strings composed of the atomic propositions or their negations.
Instead, an atomic proposition with value $\bot$ is not included in the set.
%
\paragraph{Linear-time temporal logic on finite traces.}
The specification of the expected system behavior is formalized using Linear-time Temporal Logic (\LTL) \cite{Pnueli;1977} over the (global) set of atomic propositions $\AP$.
We refer to the set of syntactically correct LTL formulas over $\mathit{AP}$ as \LTL. 
We assume the reader is familiar with LTL and its operators (Globally (\Globally), Eventually (\Eventually), strong Until (\Until), \ldots).
We denote by $\vDash$ the usual semantic relation between traces and formulas. 
We say that two formulas $\varphi_1$ and $\varphi_2$ are semantically equivalent if for any $w \in \Sigma^\infty$, $w \vDash \varphi_1$ iff $w \vDash \varphi_2$ and we denote this by $\varphi_1 \equiv \varphi_2$. 
%
%
In this paper, we use a finite-trace semantics (from~\cite{BauerLS11}).
A finite trace $u$ evaluates to $\top$ (resp. $\bot$) for $\varphi$ if all its infinite extensions satisfy (resp. do not satisfy) $\varphi$.
We denote this by $u \in good(\varphi)$ (resp. $u \in bad(\varphi)$). 
In monitoring and evaluating formulas, we will need to refer to the atomic propositions that have not been evaluated yet in a formula: $\apFormula(\varphi) \subseteq AP$ is the set of free atomic propositions occurring on $\varphi$.
%

%
\paragraph{Normal forms.}
To use the definition of literals, monomials and normal form with Linear-time temporal logic, we extend them to cover temporal operators:
	A \emph{literal} is an atomic proposition or the negation of an atomic proposition.
	A \emph{monomial} is a conjunction of literals and/or of temporal operators applied to any formulas.
	A formula is in \emph{normal form} if it only contains the operators $\vee, \wedge$ and $\neg$ and/or temporal operators (\Next, \Globally, \Eventually, \Until, \Release) and if negations that are not below a temporal operators are only applied to atomic propositions.
Finally, a formula is in \emph{disjunctive normal form (DNF)} if it is a disjunction of monomials.
We say that a formula is in \emph{temporal disjunctive normal form (TDNF)} if it is in DNF and each monomial is of the form $\varphi_1 \wedge \Next \varphi_2$ where $\varphi_1$ only contains propositional logic operators (i.e. it represents a condition on the present).

\paragraph{\texttt{map}, \texttt{fold}, and \texttt{filter}.}
We shall make use of functions \texttt{map}, \texttt{fold} and \texttt{filter}.
Consider two arbitrary types/sets $A$ and $B$.
Function \texttt{map} takes as argument a function $f : A \rightarrow B$ and a set $S$ containing elements of some type $A$.
It then returns the set $S' = \{ f(s) \mid s \in S\}$.
Function \texttt{filter} takes as argument a predicate $p$ over elements of $A$ and a set $S$ of elements from $A$ and returns the subset of $S$ with elements that satisfy $p$.
Function \texttt{fold} takes as arguments a commutative function $f : A \times B \rightarrow B$, a set $S$ containing elements of type $A$ and an element $b$ of type $B$.
It is inductively defined: if $S \neq \emptyset$, it returns $\texttt{fold}(f, S \setminus s, f(s, b))$ (where $s \in S$).
Otherwise, it returns $b$.
\section{Decentralized Runtime Enforcement}
\label{sec:re}
%
%
In this section, we first define the decentralized runtime problem, stating our assumptions.
Then, we define the requirements on decentralized enforcers.
\paragraph{Problem statement and assumptions.}
Let $i_k \in \Sigma_k^*$ (with $k \in [1 \ldots n]$) be the local trace of each component $C_k$ and $i \in \Sigma^*$ the global trace obtained from the union of the local traces.
For $k \in [1 \ldots n]$, local trace $i_k$ is input to enforcer $M_k$.
Similarly, let $o_k \in \Sigma_k^*$ be the local output trace of each enforcer and $o \in \Sigma^*$ the global output trace obtained from the union of the local outputs. 

Our assumptions on the system are as follows:
\begin{itemize}
	\itemsep 0em
	\item The formula formalizing the specification is not equivalent to $\bot$.
	\item The system cannot emit a new event until the previous one has been treated by the enforcer.
	\item An enforcer $M_k$ can only read and modify the atomic propositions in $\AP_k$.
	\item All enforcers are capable to communicate with one another by exchanging messages.
	\item All exchanged messages are delivered reliably, in order, and with no alteration.
	\item Enforcers are not malicious, i.e., they do not exchange wrong information.
\end{itemize}
Intuitively, every time a new event $\sigma$ is emitted by the system, the enforcers compute the set of events that respect the specification using their local observations of $\sigma$.
They will then choose to emit one of these events and modify their local observations accordingly.
We denote this event by $E(\sigma)$.
At time $t$, we have $E(i) = E(i(1)) \cdot E(i(2)) \dotsb E(i(t)) = o$ and $E(i_k) = E(i_k(1)) \cdot E(i_k(2)) \dotsb E(i_k(t)) = o_k$.
Let $\varphi$ be the formula representing the specification to be enforced of the system and $AP$ the set of atomic propositions present in $\varphi$.
We want to obtain online decentralized enforcers so that if $\sigma \in bad(\varphi)$, then $E(\sigma) \notin bad(\varphi)$.
\begin{example}[Running example]
We will illustrate each part of the subsequent enforcement algorithms on formula $\phi = \neg (\Globally a \vee \Eventually b)$.
We consider an example system with two components $C_1$ and $C_2$.
We use two enforcers $M_1$ and $M_2$ with $\AP_1 = \{a\}$ and $\AP_2 = \{b\}$.
The initial event emitted by the system is $\sigma = \{a\}$ and the enforcer $M_1$ is doing the initialization (this is an arbitrary choice).
\end{example}
For the remainder of this paper, unless specified otherwise, $\sigma \in \Sigma$ represents the global event emitted by the system, $E(\sigma) \in \Sigma$ denotes the event outputted by the enforcers and $\varphi \in LTL$ represents the formula to be enforced.
At timestamp 1, $\varphi$ is equal to the specification formula $\varphi_{init}$ and then, at timestamp $t+1$, $\varphi$ is equal to $\varphi^{t+1}$, the formula obtained at the end of the $t$-th timestamp. 
\paragraph{Requirements on enforcers.}
We define the requirements on a decentralized enforcer $E$.
\begin{definition}[Soundness]
	$ \forall i \in \Sigma^*,\ E(i) \notin bad(\varphi) $.
\end{definition}
An enforcer is \emph{sound} if its output is not a bad prefix of the specification.
\begin{definition}[Transparency]
	$ \forall i \in \Sigma^*, \forall \sigma \in \Sigma,\ E(i) \cdot \sigma \notin bad(\varphi) \Rightarrow E(i \cdot \sigma) = E(i) \cdot \sigma $.
\end{definition}
An enforcer is \emph{transparent} if its input is modified only when it leads to a violation of the specification.
We also define the notion of \emph{optimality}.
\begin{definition}[Optimality]
\[
\begin{array}{l}
{\forall i \in \Sigma^*},\ {\forall \sigma \in \Sigma},\ {\exists \sigma' \in \Sigma},\\
\qquad\qquad {E(i \cdot \sigma)  = E(i) \cdot \sigma'} \\
 \quad \qquad \wedge\ \forall \sigma'' \in \Sigma,\ \distance(\sigma, \sigma'', AP) < \distance(\sigma, \sigma', AP) \ \implies\ E(i) \cdot \sigma'' \in bad(\varphi).
\end{array}
\]
\end{definition}
An enforcer is \emph{optimal} if it outputs the closest event to the input that respects the property.\\

\section{Normalizing LTL Formulas}
\label{normalizing}
%
%
We transform the formula into its Temporal Disjunctive Normal Form (TDNF) to get a disjunction of monomials.
For this, we start by separating present and future obligations in the formula (Sec.~\ref{norm}) and then use an algorithm  to transform the result of the previous step into its DNF (Sec.~\ref{dnf}).
These two operations applied one after the other on the input formula yield the TDNF, in which each monomial represents a logical model of the formula and is the conjunction of two sub-formulas: a state formula (i.e. the present obligation) and a conjunction of temporal operators (i.e. the future obligation).
%
\subsection{Separating Present and Future Obligations}\label{norm}
%
The enforcers determine whether or not there is a violation of the set of properties by evaluating the corresponding formula with $\sigma$.
To achieve this, we use the expansion laws (as defined in~\cite{Pnueli;1977}) to separate what $\sigma$ needs to satisfy in the current timestamp, i.e. the present obligations, from what needs to be satisfied in the future, i.e. the future obligations. 
For example, to evaluate $\Globally a$, we first need to rewrite it as $a \wedge \Next (\Globally a)$. 
We can see that, after rewriting, $a$ has to hold on the current event $\sigma$ and that $\Globally a$ has to hold in the future.
\begin{definition}[Expansion function (\rwT)]
	Let $\varphi, \varphi_1, \varphi_2 \in \LTLm$. Function \rwT $: \LTLm \rightarrow \LTLm$ is inductively defined as follows:
	\begin{align*}
		\rwT(p) & = p,\ \text{for p} \in AP & 	\rwT(\top) & = \top \\
		\rwT(\varphi_1 \vee \varphi_2) & = \rwT(\varphi_1) \vee \rwT(\varphi_2)  &	\rwT(\bot) & = \bot \\
		\rwT(\varphi_1 \wedge \varphi_2) & = \rwT(\varphi_1) \wedge \rwT(\varphi_2) & \rwT(\neg \varphi) & = \neg \rwT(\varphi)	\\
		\rwT(\varphi_1 \Rightarrow \varphi_2) & = \rwT(\varphi_1) \Rightarrow \rwT(\varphi_2) &	\rwT(\Next  \varphi) & = \Next  \varphi \\
		\rwT(\varphi_1 \Until \varphi_2) & = \rwT(\varphi_2) \vee (\rwT(\varphi_1) \wedge \Next  (\varphi_1 \Until \varphi_2)) & \rwT(\Globally \varphi) & = \rwT(\varphi) \wedge \Next  (\Globally \varphi) \\
		\rwT(\varphi_1 \Release \varphi_2) & = \rwT(\varphi_2) \wedge (\rwT(\varphi_1) \vee \Next  (\varphi_1 \Release \varphi_2)) & \rwT(\Eventually \varphi) & = \rwT(\varphi) \vee \Next  (\Eventually \varphi) \\ 
	\end{align*}
\end{definition}
\begin{example}[\rwT]
	Recall that $\phi = \neg (\Globally a \vee \Eventually b)$.
	We have \rwT($\phi$) = $\neg (a \wedge \Next  (\Globally a) \vee b \vee \Next  (\Eventually b)) = \phi'$.
\end{example}
Any formula outputted by function \rwT\ is semantically equivalent to the input formula:
\begin{property}
\label{propRwtequiv}
$\forall \varphi \in \LTLm,\ \varphi \equiv \rwT(\varphi)$.
\end{property}
Moreover, thanks to the expansion laws, the formulas produced by \rwT\ satisfy the following syntactic property.
\begin{property}
\label{propRwt}
	Let $\varphi \in \LTLm$.
	In the syntactic tree of $\rwT(\varphi)$, any temporal operator different from \Next\ is below a $\Next $.
\end{property}
%
\subsection{Transforming to Temporal Disjunctive Normal Form}
\label{dnf}
%
A problem that arises now is knowing which formula has to be evaluated in the future based on the current observation. 
For example, with $a \Until b$, the formula to evaluate in the future will be either $\top$ or $a \Until b$ depending on the values of $a$ and $b$ in the current event.

Each monomial of a formula in DNF represents one of its model.
The transformation to DNF gives, for each model, a formula of the form $\varphi_1 \wedge \Next  \varphi_2$ where $\varphi_1$ is a conjunction of literals and $\varphi_2$ is a conjunction of temporal operators, that is, $\varphi_1$ (resp. $\varphi_2$) represents the present obligation (resp. future obligation).
Intuitively, if the present obligations of a monomial hold, then the corresponding future obligations should hold on the trace later on and should therefore be included in the formula that needs to be evaluated during the next timestamp.

In the following, we use a function \DNF\ that transforms any formula in DNF.
The details of each step of the transformation can be found in~\cite{journals/corr/abs-2107-06084}.
\begin{example}[Transformation to DNF]
	We transform $\phi'$ into its DNF:
	\[
	\DNF(\phi') = \neg a \wedge \neg b \wedge \Next (\Globally \neg b) \vee \Next (\Eventually \neg a) \wedge \neg b \wedge \Next (\Globally \neg b).
	\]
	We denote $\DNF(\phi')$ by $\phi_{DNF}$.
\end{example}
Any formula outputted by function \DNF\ is semantically equivalent to the input formula:
\begin{property}
\label{propDNFequiv}
	 $\forall \varphi \in \LTLm,\ \varphi \equiv \DNF(\varphi)$.
\end{property}
Moreover, the transformation to DNF ensures the following syntactic property:
\begin{property} \label{propDNF}
	Let $\varphi$ be an \LTL\ formula that has been rewritten by \rwT.
	In the syntactic tree of $\DNF(\varphi)$, any temporal operator different from \Next\ is below a $\Next$.
\end{property}
\begin{corollary}
	A formula rewritten by \rwT\ and then transformed into its DNF with \DNF\ is in TDNF.
\end{corollary}

%
%
\begin{definition}[Present obligation]
	$\LTLp$ is the set of formulas representing the present obligations.
	It is defined by the following grammar (where $p\in \AP$):
		$\varphi_p \in \LTLp::=  \neg p \gOR p \gOR \top \gOR \bot \gOR \varphi_p \wedge \varphi_p$.
\end{definition}
\begin{definition}[Future obligation]
	$\LTLf$ is the set of formulas representing the future obligations.
	It is defined by the following grammar (where $\varphi \in \mathit{\LTLm}$):
		$\varphi_f \in \LTLf:= \Next  \varphi \gOR \varphi_f \wedge \varphi_f$.
\end{definition}
\section{Temporal Enforcement Encoding}
\label{encoding}
%
%
Normalization (Sec.~\ref{normalizing}) provides a clear separation between present and future and ensures that the specification formula is in TDNF.
We define in Sec.~\ref{top} an encoding that associates each monomial of the formula with a pair (\textit{present}, \textit{future}) where \textit{present} (resp. \textit{future}) is a formula that represents the present obligations (resp. future obligations). Then, we define the partial function representing the state of an enforcer in Sec.~\ref{tcl}.
%
\subsection{Encoding Present and Future: Temporal Obligation Pairs (TOP)}\label{top}
%
The enforcers are now able to rewrite and evaluate the present obligations to determine whether or not $\sigma$ leads to a violation of the specification.
To represent the whole formula, we generate a set containing a pair for each monomial of the formula.
We refer to these pairs as \emph{temporal obligation pairs} (TOP). 
As the set represents a disjunction of monomials, there is a violation iff the present obligations of every pair in the set evaluate to $\bot$.

This separation allows the enforcers to only work on the present obligations.
This is useful because, for example, at timestamp $t$, the next formula to be enforced $\varphi^{t+1}$ contains some of the future obligations of $\varphi^t$ but not necessarily all of them.
Let $\mathit{present} \wedge \mathit{future}$ be a monomial in $\varphi^t$. If $E(\sigma) \vDash \mathit{present}$, then $\mathit{future}$ is included in $\varphi^{t+1}$.
Evaluating future obligations is useless if they are not included in the next formula to be enforced.
To illustrate this, consider formula $a \Until b$, we have $\rwT(a \Until b) = b \vee (a \wedge \Next (a \Until b))$. We can see here that if $b \in \sigma$ ($b$ holds), then the trace should satisfy $\top$ in the future and that, if only $a$ is true, then the trace should satisfy $a \Until b$ in the future. 
Therefore, we associate $a \Until b$ with $\{ (b, \top), (a, a \Until b)\}$.
Furthermore, splitting the formula to be enforced into smaller formulas reduces the cost of the simplification because we do not need to rewrite the future obligations in the current timestamp as stated above.

We define function \encode\ which transforms an \LTL\ formula into a set of pairs (\textit{present}, \textit{future}).
We assume that the input formula has already been rewritten by the expansion function and then transformed into its TDNF.
We know from Property \ref{propDNF} that any temporal operators is below a $\Next$.
Therefore, we do not need to define the following function for temporal operators as there cannot be any in present obligations.
\begin{definition}[Temporal obligation pair]
	Let $\varphi, \varphi_1, \varphi_2 \in \LTLm$.
	Function $\encode: \LTLm \rightarrow 2^{\LTLp \times \LTLf}$ is defined as follows:
	\begin{align*}
		\encode(\varphi_1 \vee \varphi_2) & = \encode(\varphi_1) \cup \encode(\varphi_2) &
		\encode(\varphi_1 \wedge \varphi_2) & = \{(p_{\varphi_1} \wedge p_{\varphi_2}, f_{\varphi_1} \wedge f_{\varphi_2})\} \text{ where } \\
		\encode(\neg p) & = \{(\neg p, \top)\}, \text{with } p \in \AP & &  \qquad \{(p_{\varphi_1}, f_{\varphi_1})\} = \encode(\varphi_1)  \\ \encode(p) & = \{(p, \top)\}, \text{with } p \in \AP & & \qquad \text{ and }  \\
		\encode(\Next \varphi) & = \{(\top, \varphi)\} & & \qquad \{(p_{\varphi_2}, f_{\varphi_2})\} = \encode(\varphi_2)
	\end{align*}
\end{definition}
%

\begin{example}[Temporal obligation pair]
	Let $\phi_{1} = \neg a \wedge \neg b \wedge \Next (\Globally \neg b)$ and $\phi_{2} = \Next (\Eventually \neg a) \wedge \neg b \wedge \Next (\Globally \neg b)$, we have $\encode(\phi_{1}) = \{(\neg a \wedge \neg b, \text{ } \Globally \neg b)\}$ and $\encode(\phi_{2}) = \{(\neg b, \text{ } \Eventually \neg a \wedge \Globally \neg b)\}$.
	As $\phi_{DNF} = \phi_1 \vee \phi_2$, we have $\encode(\phi_{DNF}) = \{(\neg a \wedge \neg b, \text{ } \Globally \neg b), (\neg b, \Eventually \neg a \wedge \Globally \neg b)\}$.
	We denote by $\phi_{TOP}$ the result of $\encode(\phi_{DNF})$.
\end{example}
\begin{property}
\label{propEncod}
	Let $S \in 2^{\LTLp \times \LTLf}$.
	We have:	$\forall \varphi \in \LTLm,\ \encode(\varphi) = S \implies \varphi \equiv \bigvee_{ (p, f) \in S} p \wedge f$.
\end{property}
Using the semantics of LTL, we obtain the following corollary.
\begin{corollary}
	Let $\sigma \in \Sigma^*$, $ \sigma \notin bad(\varphi)$ iff \ $\exists (p, f) \in \encode(\varphi), \sigma \notin bad(p) \wedge (sigma \notin bad(f)$.
\end{corollary}
%
\subsection{Temporal Correction Log (TCL, Enforcer State)}
\label{tcl}
%
To enforce the specification, we explore the correction events $\sigma'$ s.t. $\sigma' \notin bad(\varphi)$.
For this, we define the Temporal Correction Log (TCL) which serves as a state of the enforcer, encoding the status of the exploration.
The TCL is a function that associates events of the alphabet with a pair containing a set of temporal obligation pairs and a natural number.
We denote the set of all possible TCL by \TCL\ and an object from this set by \tcl.
When the event observed by the enforcers is $\sigma$ and $\tcl(\sigma') = (S, n)$, it means that the LTL formula that would need to be satisfied if the enforcers choose to produce event $\sigma'$ as output is encoded by the set of TOP $S$ and that the distance between $\sigma$ and $\sigma'$ is $n$.
A TCL is built incrementally: each enforcer updates the values associated with the events using their local observations.
It is a partial function $\tcl: \Sigma \rightarrow 2^{LTL_p \times LTL_f} \times \mathbb{N}$.
Initially, the state of the enforcers is initialized to $[ \emptyset \mapsto (\encode(\DNF(\rwT(\varphi))), 0)]$.
%
%
\section{Evaluating the Formula}
\label{evaluation}
%
%
After initializing their state, the enforcers have to evaluate the formula to be able to choose the output event.
To achieve this, they update their state using their local observations and send it to other enforcers to gather information on the global event.
In Sec.~\ref{update}, we define how the state of an enforcer is updated using its local observation and we then give in Sec.~\ref{reduction} a function to reduce the size of the state of an enforcer as well as defining the communication between the enforcers. 
%
\subsection{Updating the Temporal Correction Log}
\label{update}
%
When an enforcer receives the state of another one (i.e. a TCL), it updates its domain by adding its local observations. 
Each new event is associated with an updated pair and the old events are removed from the domain.
The set of TOP is updated by rewriting the present obligations and the distance metric is updated using the new local observation.

We define  the local function used by each enforcer for the rewriting of the present obligations using a local observation $\sigma'' \in \Sigma_i$.
This function uses the set of local atomic propositions to differentiate an atomic proposition that does not hold on $\sigma''$ from one that has not been observed yet.
\begin{definition}[Rewriting of the present obligations]
Let $\varphi, \varphi_1, \varphi_2 \in \LTLp$ and $\sigma'' \in 2^{\AP_i}$.
On enforcer $M_i$, function \rwi $: \LTLp \times \Sigma_i \rightarrow \LTLp$ is defined as: 
\begin{align*}
	\rwi(p \in AP, \sigma'') & =
	\begin{cases}
		\top & \text{if } p \in \sigma'' \\
		\bot & \text{if } p \notin \sigma'' \wedge p \in AP_i \\
		p & \text{otherwise}
	\end{cases} &
	\rwi(\neg \varphi, \sigma'') & = \neg \rwi(\varphi, \sigma'') \\
	\rwi(\varphi_1 \vee \varphi_2, \sigma'') & = \rwi(\varphi_1, \sigma'') \vee \rwi(\varphi_2, \sigma'') &	\rwi(\varphi_1 \wedge \varphi_2, \sigma'') & = \rwi(\varphi_1, \sigma'') \wedge \rwi(\varphi_2, \sigma'') 
\end{align*}
\end{definition}
\textbf{Note:} These cases are sufficient because we only use function \rwi\ on present obligations. Therefore, there is no temporal operators (nor implications or equivalences thanks to function \DNF). \\

We now build the set of atomic propositions that still need to be evaluated (the remaining atomic propositions in the formula).
When this set is empty, every atomic proposition of the formula has been evaluated, which means the evaluation phase is over and we then need to decide about the event to emit.
Therefore, we define function $\apr: \TCL \rightarrow 2^{\AP}$, which builds this set using \apFormula.
More precisely, \apr\ yields the union of the result of \apFormula\ on the present obligations of every set of TOP in the codomain of the \tcl, that is, $\apr(\tcl) = \texttt{fold}(\cup, \texttt{ map}(\apFormula, \{p \mid (p, f) \in \texttt{top}, (\texttt{top}, -) \in \mathit{codom}(\tcl)\}), \ \emptyset)$.

Additionally, to prevent useless rewritings with atomic propositions that are not in the formula, we update the \tcl\ using each event from $2^{AP_i \cap \apr(\tcl)}$ instead of $\Sigma_i$.
For example, in formula $\Eventually a$, we have $\apr(\tcl) = \{a\}$. 
Now suppose an enforcer $M$ with $AP_i = \{a, b\}$ has to update that formula. 
$M$ can observe four events locally: $\emptyset, \{a\}, \{b\}$ and $\{a, b\}$. 
However, as $b$ is not in the formula, $\emptyset$ and $\{b\}$ both yield the same result after rewriting (likewise with $\{a\}$ and $\{a, b\}$). 
Therefore, we can see that evaluating the formula using atomic propositions that are not present in it leads to "useless" rewriting because the result is the same with at least one other local observation.
This also means that each enforcer only receives the TCL once during the evaluation as all the atomic propositions they can observe locally are replaced by either $\top$ or $\bot$.
In the remainder of this paper, we denote $2^{AP_i \cap \apr(\tcl)}$ by $\Sigma_i^r$ which represents, intuitively, the set of all local observations of enforcer $M_i$ containing only atomic propositions that are present in the formula.
\begin{example}[Rewriting of the present obligations]
	Recall that $\phi_{TOP} = \{(\neg a \wedge \neg b, \Globally \neg b), (\neg b, \Eventually \neg a \wedge \Globally \neg b)\}$. 
	After initialization, we have dom(\tcl) = $\{\emptyset\}$ and \tcl($\emptyset$) = ($\phi_{TOP}$, 0).
	Here, the present obligation of the first monomial contains $a$ and $b$, the ones from the second monomial only contains $b$. 
	Therefore, $\apr(\tcl) = \apFormula(\neg a \wedge \neg b) \cup \apFormula(\neg b) = \{a, b\}$.
	We have $AP_1 = \{a\}$ so $\Sigma_1 = \{ \emptyset, \{a\} \} = \Sigma_1^r$. 
	Evaluating the present obligations using the local observations of $M_1$ yields the following formulas:
	\begin{align*}
	\rwi(\neg a \wedge \neg b, \emptyset) = \neg b &&
	\rwi(\neg b, \emptyset) = \neg b &&
	\rwi(\neg a \wedge \neg b, \{a\}) = \bot &&
	\rwi(\neg b, \{a\}) = \neg b
\end{align*} 
\end{example}

\begin{wrapfigure}{L}{0.5\textwidth}
	\vspace{-\abovedisplayskip}
	\begin{minipage}{0.5\textwidth}
		\begin{algorithm}[H]
			\caption{Update of the current enforcer state: \updateTCL(\tcl, $\sigma$, \apr(\tcl))}
			\begin{algorithmic}[1]
				\FOR{each $\sigma' \in dom(\tcl)$}
				\STATE Let $(\mathit{pf}, n)$ = $\tcl(\sigma')$
				\STATE $dom(\tcl) = dom(\tcl) \setminus \sigma'$
				\FOR{each $\sigma'' \in \Sigma_i^r$}
				\STATE Let $\mathit{pf'} = \texttt{map}(\lambda(p, -) . \rwi(p, \sigma''), \ \mathit{pf})$
				\STATE $\tcl = \tcl[\sigma'' \cup \sigma' \rightarrow \ (\mathit{pf'}, \  n + \distance(\sigma'', \sigma, AP_i \cap \apr(\tcl)))]$
				\ENDFOR
				\ENDFOR
			\end{algorithmic}
		\end{algorithm}
	\end{minipage}
\end{wrapfigure}
We now define function \updateTCL\ $: \TCL \times \Sigma_i \times AP \rightarrow \TCL$ used by the enforcers to update their state.
The function takes three arguments: the state of enforcer to update, its local observation of the global event and the set of atomic propositions that have not been evaluated yet in any present obligation.
It is updated using every $\sigma'' \in \Sigma_i^r$.
%
The domain becomes the set $\sigma' \cup \sigma''$ (with $\sigma' \in dom(\tcl)$, an event of the "old" domain).
Let $pr$ be some present obligation. 
$pr$ is rewritten as follows: $pr$ = \rwi($pr$, $\sigma''$). 
The distance is updated: it is increased by 1 for each atomic proposition $p \in \sigma''$ that has a different truth value compared to $\sigma$, that is, if the metric was equal to $n$ before the update, it becomes $n + \distance(\sigma, \sigma'', AP_i \cap \apr(\tcl)))$.
The definition of \updateTCL\ is given in Algorithm 1.
%
\begin{example}[\updateTCL]
	Recall that $AP_1 = \{a\}$, $\apr(\tcl) = \{a, b\}$ and $\sigma = \{a\}$. Before $M_1$ updates \tcl, we have $\tcl(\emptyset) = (\phi_{TOP}, 0) = (\{(\neg a \wedge \neg b, \  \Globally \neg b), (\neg b, \Eventually \neg a \wedge \Globally \neg b)\}, \  0)$. 
	After updating \tcl\ with \updateTCL, we have:
	\begin{align*}
		\tcl(\emptyset) = (\{ (\neg b, \Globally \neg b), (\neg b, \Eventually \neg a \wedge \Globally \neg b) \}, \  1) &&
		\tcl(\{a\}) = (\{ (\bot, \Globally \neg b), (\neg b, \Eventually \neg a \wedge \Globally \neg b) \}, \  0)
	\end{align*}
\end{example}
%
\subsection{Reduction of the Obligations Set and Communication}\label{reduction}

  \begin{wrapfigure}{R}{0.54\textwidth}
	\begin{minipage}[t][4.8cm]{0.54\textwidth}
	\vspace{-1em}
		\begin{algorithm}[H]
			\caption{Reduce the size of the domain of the TCL}
			\begin{algorithmic}[1]
				\FOR{each $\sigma' \in dom(\tcl)$}
				\STATE Let $(\mathit{pf}, -) = \tcl(\sigma')$
				\STATE $\mathit{pf}' = \texttt{filter}((\lambda(p,f).(p \not\equiv \bot \wedge f \not\equiv \bot)), \ \mathit{pf})$
				\IF{$\mathit{pf}' == \emptyset$}
				\STATE $dom(\tcl) = dom(\tcl) \setminus \sigma'$
				\ELSE
				\STATE $\tcl = \tcl[\sigma' \rightarrow (\mathit{pf}', \  n)]$
				\ENDIF
				\ENDFOR	
			\end{algorithmic}
		\end{algorithm}
	\end{minipage}
\end{wrapfigure}
An issue induced by Algorithm 1 is that the size of the domain of the \tcl\ doubles for each atomic propositions in $AP_i \cap \apr(\tcl)$. 
Therefore, we reduce its size by removing certain elements that become useless after rewriting. 
First, we reduce the size of the elements of the codomain (the images) by removing the pairs in which either the present or the future obligations have been evaluated to $\bot$, i.e. the monomials evaluated to $\bot$. 
Then, we remove from the domain the events that are associated with a pair containing an empty set of TOP as these events do not satisfy any present obligations. 
The second reduction guarantees that any event in the domain at the end of the evaluation does not lead to a violation of $\varphi$.
Therefore, this also guarantees that adding the emitted event to the trace will not form a bad prefix of the formula.

For this, we define \reduction$: \TCL \rightarrow \TCL$ which implements the two aforementioned reductions in Algorithm~2.

%
\begin{example}[\reduction]
	In example 5, we got $\tcl(\{a\}) = (\{ (\bot, \Globally \neg b), (\neg b, \Eventually \neg a \wedge \Globally \neg b) \}, \  0)$.
	The set of obligations associated with this event contains a monomial in which the present obligations have been evaluated to $\bot$. 
	Therefore, $\tcl(\{a\}) = (\{ (\neg b, \Eventually \neg a \wedge \Globally \neg b) \}, \  0)$.
\end{example}
After this reduction, if there still is at least one atomic proposition to evaluate, then the current enforcer communicates its state to another enforcer that can update it, i.e. an enforcer that can locally observe one of the remaining atomic propositions in any of the present obligations.
If there are multiple enforcers that can rewrite the formula, the one with the smallest index is chosen.
If the formula has been evaluated in its entirety, the decision rule can be applied to choose the event to emit.
\begin{example}[Communication]
	We can see in the previous example that $b$ has not been evaluated yet so the \tcl\ is sent to a enforcer that can observe this atomic proposition: $M_2$, in that case.
\end{example}
%
\section{Enforcement using a Decision based on Global Exploration}
\label{algoI}
%
%
We now define the decision rule applied by the last enforcer in Sec.~\ref{decision1} and the algorithm itself in Sec.~\ref{algo1}. We state some properties of the algorithm in Sec.~\ref{prop1}.
%
\subsection{Decision Rule}\label{decision1}
%
The decision rule is used by the enforcers to determine the emitted event. 
Let $\tcl_f$ be the partial function representing the final state of the enforcer. 
Once an event has been chosen, each enforcer will modify its local observations accordingly (if needed). 
The last enforcer that updates the \tcl\ needs to send its state to all the other enforcers so that they all apply the decision rule and choose an event in parallel. \\

To respect \emph{transparency}, we simply choose the event that has the least number of changes compared to $\sigma$. 
We know (from section \ref{reduction}), that the events $\sigma'$ so that $u \cdot \sigma' \in bad(\varphi)$ (with $u \in \Sigma^*$ the trace up until the current timestamp and $\varphi \in LTL$ the property to enforce) have been removed from the domain of $\tcl_f$. 
Therefore, if $u \cdot \sigma \notin bad(\varphi)$, then $\sigma$ is the only element of the domain with its distance equal to 0 so $\sigma$ will be emitted. 
If $u \cdot \sigma \in bad(\varphi)$, it is possible to have multiple events with the same distance.
Let $\tcl_{\mathrm{candidates}}$ be the set of events with the least number of changes from $\sigma$: $\tcl_{\mathrm{candidates}} =  \{ \sigma' \in \tcl_f \ |\ \tcl_f(\sigma') = (-, n_{\mathrm{min}})\}$ so that $n_{\mathrm{min}} = min(\{n \ |\ (-, n) \in \mathit{codom}(\tcl_f)\})$.
%
If $\sigma'$ is chosen, then the local event emitted by each enforcer $M_i$ is $\sigma' \cap AP_i$.

If we choose to emit event $\sigma'$, then $\varphi^{t+1}$ is the disjunction of the future obligations associated with the emitted event and we have $\varphi^{t+1} = \texttt{fold}(\ (\lambda x,(-, f).(x \vee f)), \  \mathit{pf}, \  \bot)$) with $(\mathit{pf}, -) = \tcl_f(\sigma')$.
Therefore, if $(\top, \top)$ is included in the set of TOP of any events in $\TCL_{\mathrm{candidates}}$, then $\varphi^{t+1} = \top$.
In this particular situation, we can stop the enforcers (any event is a good prefix of $\top$) so, if there exists $\sigma' \in \tcl_{\mathrm{candidates}}$ so that $\TCL(\sigma') = (\mathit{pf}, -) \wedge (\top, \top) \in \mathit{pf}$ then we reduce $\TCL_{\mathrm{candidates}}$ so that it only contains these $\sigma'$.
Finally, if we still have $\#\TCL_{\mathrm{candidates}} > 1$, we choose the event to emit from this set arbitrarily (but deterministically).
Every enforcer applies the decision rule but the choice of the verdict is deterministic so they will all choose the same one. 
This implies that $\varphi^{t+1}$ is the same for all of them which means that they all know the next formula to enforce without any additional communication.
\begin{example}[Decision rule]
	After the update of $M_2$, we have:
	\begin{align*}
		\tcl(\emptyset)= (\{ (\top, \top), (\top, \Eventually \neg a \wedge \Globally\neg b) \}, \  1 ) &&
		\tcl(\{a\})=(\{ (\top, \Eventually \neg a \wedge \Globally\neg b) \}, \  0)
	\end{align*}
	The events $\{a, b\}$ and $\{b\}$ have been removed as they lead to a violation of $\phi$ ($\Globally\neg b$ evaluates to $\bot$ if $b = \top$).
	We apply the decision rule to choose the event to emit.
	Here, $\sigma$ is chosen because it does not lead to a violation so we do not need to modify it.
	The local event emitted by $M_1$ (resp. $M_2$) is $\sigma'' = \{a\} \in \Sigma_1$ (resp. $\sigma'' = \emptyset \in \Sigma_2$).
	The formula that needs to be monitored during the next timestamp is $\phi^{t+1} = \Eventually \neg a \wedge \Globally\neg b$.
\end{example}
%
\subsection{Enforcement Algorithm}\label{algo1}
%
Let $\mathcal{M} = \{M_1, \ldots, M_n\}$ be the set of enforcers. 
Algorithm~\ref{algo:enforcement:global} is a local algorithm run by each enforcer.
Enforcer $M_1$ is chosen arbitrarily to be the one initializing the evaluation.
$M_1$ also updates its state immediately if it is able to, i.e. if the formula contains an atomic propositions in $AP_1$.
All the other enforcers start by waiting to receive a \tcl.
Each enforcer $M_j$ takes its local observation of the global event $\sigma \in \Sigma$ emitted by the system as input.
Once the computation is over, they output their local observations of the event $\sigma' \in \Sigma$, that is different from $\sigma$ iff $u \cdot \sigma \in bad(\varphi)$, where $u$ is the trace up until the current timestamp t, $u \in \Sigma^*$.
The enforcer also builds the formula $\varphi^{t+1}$ based on the emitted event. 
\begin{algorithm}
\caption{Enforcement on $M_j$ using a decision based on global exploration}
\label{algo:enforcement:global}
\begin{algorithmic}[1]
	\STATE \textbf{Initialization: } If $j == 1$, \tcl\ is initialized to $[\emptyset \mapsto (\encode(\DNF(\rwT(\varphi^t))), 0)]$. \\
	Otherwise, wait until a \tcl\ is received from another enforcer. If $\apr(\tcl) = \emptyset$ in the received \tcl, jump to step 5.
	\STATE \textbf{Evaluation: } Compute \apr(\tcl), the set of atomic propositions that have not been evaluated yet. 
	Update \tcl\ using \updateTCL(\tcl, $\sigma$, $AP_j \cap \apr(\tcl)$). 
	\STATE \textbf{Reduction: } Reduce the domain of \tcl\ by removing the monomials that evaluate to $\bot$ and the bad prefixes of $\varphi^t$ using \reduction(\tcl).
	\STATE \textbf{Communication: } If $\apr(\tcl) \neq \emptyset$, let $\mathcal{M}' \subseteq \mathcal{M}$ be the set of enforcers $M_k$ so that $AP_k \cap \apr(\tcl) \neq \emptyset \wedge j \neq k$.
	\tcl\ is sent to enforcer $M_{k_{\mathrm{min}}}$ with $k_{\mathrm{min}} = \mathit{min} \{k \mid M_k \in \mathcal{M}'\}$.
	Wait until a \tcl\ is received from another enforcer. \\
	Otherwise, send \tcl\ to all the other enforcers so that they can apply the decision rule. 
	\STATE  \textbf{Decision: } Let $\tcl_f$ be the final state of \tcl. Apply the decision rule to choose the event $\sigma'$ to emit and set $\varphi^{t+1}$ to $\texttt{fold}(\text{ }(\lambda x, (-, f).(x \vee f)), \ \mathit{pf}, \ \bot)$, with $(\mathit{pf}, -) = \tcl_f(\sigma')$.
\end{algorithmic}
\end{algorithm}
%
\subsection{Properties}
\label{prop1}
%
The number of messages sent is bounded: in the worst case scenario, that is, if the formula contains at least one local observation of each enforcer, the \tcl\ is sent to every enforcer during the evaluation and once more at the end so that every enforcer can apply the decision rule.
Therefore, if we denote by $\delta_{\mathrm{init}}$ the time needed to initialize the state of the enforcer (Algorithm 2), by $\delta_{\mathrm{update}}$ the time needed for one update of the \tcl\ (update with each element of $2^{AP_i \cap \apr(\tcl)}$) and by $\delta_{\mathrm{decision}}$ the time it takes to apply the decision rule.
The delay between two events emitted by the system is, at worst, $\#\mathcal{M} \times \delta_{\mathrm{update}} + \delta_{\mathrm{init}} + \delta_{\mathrm{decision}}$ (every enforcer executes the decision rule at the same time).
The size of the messages is also bounded: as we rewrite the obligations using every possible events in $\Sigma$, the size of the domain doubles for each observed atomic proposition.
Therefore, denoting by $n_{\AP}$ the number of atomic propositions in the formula, the size of the domain is bounded by $2^{n_{\AP}}$ elements and the size of the last message sent can be at most $2^{n_{\AP} - 1}$.
At worst, the domain contains $2^{\#\AP}$ elements at the end.
It is worth noting that the transformation to TDNF is costly (exponential in the size of the formula) and that, in the worst case scenario, if \reduction\ does not remove elements from the domain, its size grows exponentially as well, which means that, in turn, $\delta_{\mathrm{update}}$ gets longer every timestamp. 
$\delta_{\mathrm{decision}}$ is negligible compared to the other terms. 
\begin{lemma}
\label{propRes}
	Let $\varphi^t$ be the property to enforce at timestamp t and $\varphi^{t+1}$ the formula built at the end of the algorithm (that will be monitored during timestamp $t+1$). If $\varphi^t \not \equiv \bot$, then $\varphi^{t+1} \not \equiv \bot$.
\end{lemma}
Since the initial property is assumed to be not equivalent to $\bot$, \lemmaref{propRes} implies that \algoref{algo:enforcement:global} cannot produce a formula that is equivalent to $\bot$.
\begin{property}
	By using \algoref{algo:enforcement:global} as a local enforcer on each component, we obtain a sound, transparent, and optimal enforcer (as described in \secref{sec:re}).
\end{property}
%
\section{Enforcement using a Decision based on Local Exploration}
\label{algoII}
%
%
In this section, we define another enforcement algorithm where each enforcer takes a local decision before sending its state to the next enforcer: instead of sending the whole \tcl, the current enforcer only sends a single entry with its image. 
This also means that the enforcers do not need to make a decision at the end of the enforcement round: the local emitted event corresponds to this local decision.
The receiving enforcer then initializes its state using the received event and applies the updates to it.
The point of this approach is to prevent the exponential growth of the domain of the \tcl.
The local decision rule is defined in Sec.~\ref{decision2} and the algorithm in Sec.~\ref{algo2}.
We compare the two versions in \secref{comparison} and give the properties of the second version in \secref{prop2}. A complete example is given in \secref{exampleTable}.
%
\subsection{Local Decision Rule}
\label{decision2}
%
To allow the enforcers to make a local decision, we redefine the decision rule.
Let $\tcl$ be the partial function representing the state of the enforcer after its update. 
\apr(\tcl) is recomputed after the evaluation.
Two cases are possible:
\begin{itemize}
	\item 
If $\apr(\tcl) = \emptyset$, the algorithm stops and the current enforcer decides which event to emit from a set $\tcl_{\rm candidates} =  \{ \sigma' \in \tcl_f \mid \tcl_f(\sigma') = (-, n_{min})\}$ so that $n_{min} = min(N)$ with $N = \{n \mid (-, n) \in \mathit{codom}(\tcl_f)\}$. \\
%
If there exists $\sigma' \in \tcl_{\rm candidates}$ so that $\tcl(\sigma') = (\mathit{pf}, -) \wedge (\top, \top) \in \mathit{pf}$, then $\tcl_{\rm candidates} = \{\sigma' \in \tcl_{\rm candidates} \mid \tcl(\sigma') = (\mathit{pf}, -) \wedge (\top, \top) \in \mathit{pf}\}$ for the same reasons as in the first algorithm: if $\varphi^{t+1} = \top$, we do not need to enforce the formula anymore as any event is a good prefix of $\top$.
Finally, if there are multiple elements in $\tcl_{\rm candidates}$, one is chosen arbitrarily (and deterministically).
	\item 
If $\apr(\tcl) \neq \emptyset$, the current enforcer decides which element of the domain it will send to the next enforcer (with its image). 
The element is chosen from the set $\tcl_c = \{ (\sigma', (\mathit{pf}, n_{min})) \ |\ \sigma' \in dom(\tcl) \wedge \tcl(\sigma') = (\mathit{pf}, n_{min})\}$ so that $n_{min} = min(N)$ with $N = \{n \mid (-, n) \in \mathit{codom}(\tcl_f)\}$, that is, it is chosen among the elements that have the smallest distance to $\sigma$. \\
If there are several events that respect the property mentioned above, the set is reduced to the events that have the most models, i.e. $\tcl_c = \{(-, (\mathit{pf}, -)) \in \tcl_c \mid \nexists  (-, (\mathit{pf}', -)) \in \tcl_c, \mathit{pf} \neq \mathit{pf}' \wedge |\mathit{pf}'| > |\mathit{pf}|\}$ (the formulas that have the highest amount of monomials).
If there are still multiple elements in $\tcl_c$, one is chosen arbitrarily (and deterministically).
\end{itemize}
Just as in the first algorithm, if $\sigma'$ is chosen, then $M_i$ outputs $\sigma' \cap AP_i$.
As only the last enforcer that applies the decision rule has sufficient information to determine $\varphi^{t+1}$, it needs to be sent to the other enforcers at the end of the evaluation.
%
\subsection{Enforcement Algorithm}
\label{algo2}
%
To obtain the enforcement algorithm in the case of decision based on local exploration, we update \algoref{algo:enforcement:global} as follows.

In step 1, the (current) enforcer waits for a \tcl\ or the formula to enforce in the next timestamp.
If it receives a formula, the execution of the algorithm for the current timestamp stops.

In step 4, the enforcer always applies the local decision rule to choose an event from the domain.
It then removes all the other events from the domain and, if $\apr(\tcl) \neq \emptyset$, it sends \tcl\ to the next enforcer using the same criteria as in the first algorithm.
Otherwise, it waits for the formula to enforce in the next timestamp instead of a \tcl.
When it receives the formula, the current timestamp stops (the enforcer does not go to step 5).

In step 5, the enforcer does not apply the decision rule as it already did it in the previous step.
It only builds the formula to enforce at the next timestamp and sends it to all the other enforcers.

%
\subsection{Comparison}
\label{comparison}
%
First, let us notice that the algorithm with the local decision rule guarantees \emph{soundness} and \emph{transparency}, but not \emph{optimality}.
Indeed, as the algorithm always sends an event associated with a set of future obligations that has at least one solution, there is always a solution.
Otherwise, this event would not be in the domain.
This guarantees soundness.
Transparency is guaranteed since the input event is removed from the domain only if it violates the property.
However, as the set of all possible events over $\AP$ is not entirely explored, the event emitted at the end might not be the best (w.r.t. the distance to $\sigma$).
Hence, \emph{optimality} is not guaranteed.

Regarding performance, despite requiring the transformation to DNF during the initialization, the algorithm presents a significant improvement regarding the size of the messages and of the domain (it prevents their exponential growth).
Furthermore, the size of the domain is much smaller, it only contains one element before the update instead of, at worst, $2^{k-1}$ after $k$ updates, which allows the enforcers to only do one update per element of $2^{\AP_i \cap \apr(\tcl)}$ instead of $\#2^{AP_i \cap \apr(\tcl)} \times 2^{k-1}$ updates (one per local observation per element of the domain).

However, the algorithm has some drawbacks: the decision rule has to be applied by each enforcer, although this is not that significant considering it is negligible compared to the other operations.
Moreover, this version does not guarantee \emph{optimality}.
It is worth noting that the first version of the algorithm is better than this one in one very specific scenario: if the domain of \tcl\ is reduced to a single element with \reduction\ at the end of every update, then this version is worse because it does not improve the size of the messages/domain but the enforcers still have to apply the local decision rule after every update instead of once in the first version. 
Even in this situation, the difference is not significant unless there is a large number of enforcers.
%
\subsection{Properties}
\label{prop2}
%
Just as in the first version, the number of messages sent is bounded: in the worst case, that is, if the formula contains at least one local observation of each enforcer, one message is sent to each enforcer.
An additional message is sent to all the enforcers (except 1) at the end to communicate $\varphi^{t+1}$.
Therefore, the delay between two events emitted by the system is, at worst, $(\delta_{\mathrm{update}} + \delta_{\mathrm{decision}}) \times \#\mathcal{M} + \delta_{\mathrm{init}}$ with $\delta_{\mathrm{init}}, \delta_{\mathrm{update}}$ and $\delta_{\mathrm{decision}}$ defined as in Sec.~\ref{prop1}.
As we only send a pair containing an element of the domain of \tcl\ and its image, the size of the message is quite small.
It is not completely constant because the size of the image may vary (although not by much).
%
\subsection{Complete Example: Decentralized Traffic Lights}\label{exampleTable}
%
\begin{table}[H]
\centering
\renewcommand{\arraystretch}{1.15}
\caption{Enforcing $\Globally((g_1 \wedge g_3 \wedge \neg (g_2 \vee  g_4)) \vee (\neg (g_1 \vee g_3) \wedge g_2 \wedge g_4))$ given event $\sigma = \{g_1, g_2, g_3\}$.}
	\begin{tabular}{rl}
		\small Present/Future & $\rwT(\varphi) = (g_1 \wedge g_3 \wedge \neg (g_2 \vee g_4) \vee \neg (g_1 \vee \neg g_3) \wedge g_2 \wedge g_4) \wedge \Next \varphi = \varphi_{rwT}$ \\
		\rowcolor{Gray}
		\small Transf. to TDNF & $\DNF(\varphi_{rwT}) = (g_1 \wedge g_3 \wedge \neg g_2 \wedge \neg g_4 \wedge \Next \varphi) \vee\ (\neg g_1 \wedge \neg g_3 \wedge g_2 \wedge g_4 \wedge \Next \varphi) = \varphi_{TDNF}$ \\
		\small Initial \tcl & $[\emptyset \mapsto \encode(\varphi_{TDNF}) = (\{ (g_1 \wedge g_3 \wedge \neg g_2 \wedge \neg g_4, \varphi), (\neg g_1 \wedge \neg g_3 \wedge g_2 \wedge g_4, \varphi) \}, \ 0) ]$ \\
		\rowcolor{Gray}
		\hline
		\small Evaluation ($M_1$) & $[ \emptyset \mapsto (\{ \textcolor{dark-gray}{(\bot, \varphi)}, (\neg g_3 \wedge g_2 \wedge g_4, \varphi) \}, \ 1 ), \{g_1\} \mapsto (\{ (g_3 \wedge \neg g_2 \wedge \neg g_4, \varphi), \textcolor{dark-gray}{(\bot, \varphi)} \}, \ 0)]$ \\
		\small Decision ($M_1$) & The entry corresponding to $\{g_1\}$ is chosen, others are removed. \\
		\rowcolor{Gray}
		\small Evaluation ($M_2$) & $[ \{g_1\} \mapsto \{ (g_3 \wedge \neg g_4, \varphi) \}, \ 1), \{g_1, g_2\} \mapsto (\{ \textcolor{dark-gray}{(\bot, \varphi)} \}, \ 0) ]$ \\
		\small Decision ($M_2$) & $\{g_1\}$ is the only event left in the domain. It is therefore chosen by default. \\
		\rowcolor{Gray}
		\small Evaluation ($M_3$) & $[ \{g_1\} \mapsto (\{ \textcolor{dark-gray}{(\bot, \varphi)} \}, \ 2),\{g_1, g_3\} \mapsto (\{ (\neg g_4, \varphi) \}, \ 1) ]$  \\
		\small Decision ($M_3$) & $\{g_1, g_3\}$ is the only event left in the domain. It is therefore chosen by default. \\
		\rowcolor{Gray}
		\small Evaluation ($M_4$) & $[ \{g_1, g_3\} \mapsto (\{ (\top, \varphi) \}, \ 1), \{g_1, g_3, g_4 \} \mapsto (\{ \textcolor{dark-gray}{(\bot, \varphi)} \}, \ 2)]$ \\
		\small Decision ($M_4$) & $\{g_1, g_3\}$ is the only event left in the domain. It is therefore chosen by default. \\
		\hline
		\rowcolor{Gray}
		\small Next formula & $\varphi^{2} = \varphi$ \\
	\end{tabular}
	\label{tab:example}
\end{table}
We consider a crossroad and its four traffic lights. 
We have $AP_i = \{g_i, y_i, r_i\}, i \in \{1, 2, 3, 4\}$.
Variable $g_i$ (resp. $y_i$ and $r_i$) indicates whether or not the green (resp. yellow and red) light of the $i$-th traffic light is on ($g_i = \top$ means it is on). 
Let $\varphi$ be the property representing the following specification: \emph{At any time, exactly two opposed traffic lights must be green at the same time}. 
We have: $\varphi = \Globally((g_1 \wedge g_3 \wedge \neg (g_2 \vee  g_4)) \vee (\neg (g_1 \vee g_3) \wedge g_2 \wedge g_4))$.
Let $M_1, M_2, M_3$ and $M_4$ be the four enforcers associated with each traffic light (one per traffic light). 
Let $\sigma \in \Sigma$ be the event emitted by the system, $\sigma = \{ g_1, g_2, g_3 \}$ which means that only the green light of the first three traffic lights is on.
All the transformations are given in Table~\ref{tab:example}.
A detailed description of this example on both enforcement algorithms is in~\cite{journals/corr/abs-2107-06084}.
In the table, shaded pairs are removed from the set through function \reduction.
Then, events associated to an empty set of TOP are removed from the domain.
\section{Conclusions and Future work}
\label{sec:conclusion}
%
\paragraph{Conclusions.}
This paper introduces the problem of \emph{decentralized runtime enforcement} for systems without a global observation/control point.
We give two decentralized enforcement algorithms for LTL formulas.
Both algorithms guarantee \emph{soundness} and \emph{transparency}. 
The first also guarantees the \emph{optimality} of the modifica\-tions done by the enforcer in terms of distance to the original event emitted by the system while the second comes with a drastically reduced cost (both in terms of time and space).
\paragraph{Future work.}
The natural extension of this work is its implementation to empirically evaluate it (on LTL specification patterns~\cite{Dwyer1999}, for example) in terms of computational and communication costs.
Then, we plan to integrate it into the THEMIS tool~\cite{elhokayem:hal-01653727} which currently only supports verification.
Our enforcement algorithms can also be extended in several ways.
First, using LTL formula rewriting has a few drawbacks and in particular, rewriting renders the analysis of the runtime behavior of monitors hard to predict as it depends on the simplification function applied to LTL formulas after rewriting.
Alternatively, we consider encoding the specification using automata, for example.
Finally, we shall also consider timed properties as they are much more expressive. 
There are some approaches for the runtime enforcement of timed properties (see~\cite{FalconeP19} for an overview), but all are centralized.
%
%
%
\paragraph{Acknowledgment.}
Y. Falcone acknowledges the support from the H2020-ECSEL-2018-IA call – Grant Agreement number 826276 (CPS4EU), the European Union’s Horizon 2020 research and innovation programme - Grant Agreement number 956123 (FOCETA), from the French ANR project ANR-20-CE39-0009 (SEVERITAS), the Auvergne-Rh\^one-Alpes research project MOAP, and LabEx PERSYVAL-Lab (ANR-11-LABX-0025-01) funded by the French program Investisse\-ment d’avenir.
%
\bibliography{biblio}
\bibliographystyle{eptcs}
\end{document}